\documentclass[aps,prb,floatfix,twocolumn,showpacs,10pt]{revtex4-1}

\usepackage{epsfig,amsmath,amssymb}

\begin{document}

\title{Electric control of tunneling energy in graphene double dots}
\author{Martin Raith$^1$, Christian Ertler$^1$, Peter Stano$^{2,3}$, Michael Wimmer$^4$, and Jaroslav Fabian$^1$}
\affiliation{$^1$Institute for Theoretical Physics, University of Regensburg, D-93040 Regensburg, Germany\\
$^2$RIKEN Center for Emergent Matter Science, 2-1 Hirosawa, Wako, Saitama, 351-0198 Japan\\
$^3$Institute of Physics, Slovak Academy of Sciences, 845 11 Bratislava, Slovakia\\
$^4$Instituut-Lorentz, Universiteit Leiden, P.O. Box 9506, 2300 RA Leiden, The Netherlands}

\vskip1.5truecm

\begin{abstract}
We theoretically investigate the spectrum of a single electron double quantum dot, defined by top gates in a graphene with a substrate induced gap. We examine the effects of electric and magnetic fields on the spectrum of localized states, focusing on the tunability of the inter-dot coupling. We find that the substrate induced gap allows for electrostatic control, with some limitations that for a fixed inter-dot distance, the inter-dot coupling can not be made arbitrarily small due to the Klein tunneling. On the other hand, the proximity of the valence band in graphene allows for new regimes, such as an $npn$ double dot, which have no counterparts in GaAs. 
\end{abstract}


\maketitle

\section{Introduction}
Graphene is an outstanding material in many respects.\cite{geim2007:NM,avouris2007:NN,castroneto2009:RMP,abergel2010:AP} The unique band structure of graphene---a linear dispersion, which can be described by the Dirac equation of massless particles\cite{castroneto2009:RMP,abergel2010:AP}---has attracted research efforts to utilize its electronic properties for novel devices and applications.\cite{lin2010:S,wu2011:N} Since the experimental discovery in 2004,\cite{novoselov2004:S,novoselov2005:PNAS} graphene research has grown remarkably. Graphene is also discussed in the context of quantum information processing,\cite{trauzettel2007:NP,guo2009:NJP,recher2010:Nano} with electrons confined in a graphene-based quantum dot. Defining gated dots by electrostatically in bulk graphene  is problematic due to the Klein tunneling.\cite{calogeracos1999:CP} The limitations can be overcome in the presence of a gap,\cite{recher2009:PRB} an energy splitting of the valence and conduction band. It arises from transverse confinement in nano ribbons with certain boundaries,\cite{trauzettel2007:NP, silvestrov2007:PRL} or from an underlying substrate.\cite{giovannetti2007:PRB,zhou2007:NM} 

Another standard technique to create quantum dots is to cut a structure with the desired geometry from a flake of graphene.\cite{geim2007:NM, stampfer2008:APL, ponomarenko2008:S, kolbl2013:U} However, so far there are no techniques for creating boundaries with atomic precision. The precise termination of the lattice, on the other hand, has qualitative consequences on the confined structure spectra:\cite{mccann2004:JPCM, akhmerov2008:PRB, vanevic2009:PRB} the gap can be induced or closed, and mid-gap, highly localized, states can arise.\cite{rozhkov2011:PR} In addition, results of theoretical models of different sophistication (DFT vs.~tight-binding vs.~continuous Dirac equation) can also differ qualitatively in these aspects.\cite{akhmerov2008:PRB,guttinger2012:RPP} Our choice to consider a gap-based dot is also to avoid all such model ambiguities.\cite{grujic2011:PRB}

Graphene single dots have been intensively investigated during the last years.\cite{trauzettel2007:NP,silvestrov2007:PRL,schnez2008:PRB,recher2009:PRB,guttinger2010:PRL,wimmer2010:PRB,abergel2010:AP,grujic2011:PRB,zarenia2011:PRB} We extend those works by theoretical investigations of the dot-dot coupling, an essential ingredient for quantum computation.\cite{bennett2000:N,loss1998:PRA} The system we consider consists of two graphene dots that are laterally coupled via a tunable barrier.\cite{dawei2013:SR} Both the dots and the barrier are defined in an infinite sheet of graphene by electric gates, with the Klein tunneling suppressed by a substrate-induced mass. 

We find that the setup we choose allows for electrostatic control over the inter-dot tunneling. Compared to GaAs dots, the most important difference is the fact that the Klein tunneling results in a minimum below which the inter-dot tunneling can not be tuned. This minimum is set by the gap and the dot geometry (the inter-dot distance) and can be further suppressed by a perpendicular magnetic field. On the other hand, the presence of the valence band states allows for interesting regimes of operation of a double dot, which have no analogs in semiconductor quantum dots in GaAs or Si.\cite{hanson2007:RMP, kloeffel2013:ARCMP}

This article is organized as follows. Sec.~\ref{model} contains the model, where we discuss the details of the electrostatic confinement potential and also comment on the numerical implementation. The numerical results and their analysis for a single dot are given in Sec.~\ref{results:singledot}. The main focus of the work is Sec.~\ref{results:doubledot}, where we analyze the tunneling energy and its electrostatic tuning. We conclude in Sec.~\ref{conclusions}.

\begin{figure}
 \centering
 \includegraphics[width=0.80\linewidth]{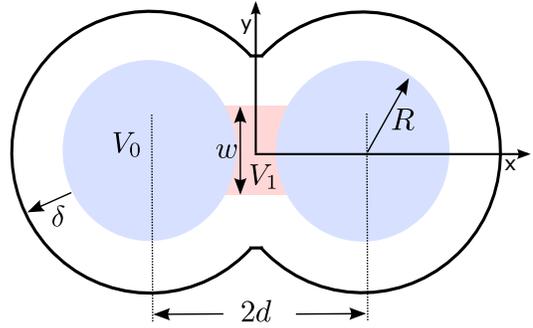}
\caption{(Color online) Illustration of the electrostatically defined graphene double dot. The top gates (colored/shaded regions) define the area of confinement, which can be adjusted for the circular dots via $V_0$ (blue areas/circles) and the bridge between the dots via $V_1$ (red area/center) separately. In numerics, we terminate the grid at a distance $\delta$ beyond the structure potential boundary, and we introduce a slightly increased mass at the edge of the grid (the outer black line).}
 \label{fig:dotsetup}
\end{figure}

\section{Model}\label{model}
We consider an infinitely large sheet of gapped graphene with a constant mass term $\Delta$. The mass, which can be induced by a substrate, e.g., boron nitride, leads to a gap of $2\Delta$, which separates the valence and conduction band at the $K$ and $K'$ points of the Brillouine zone.\cite{giovannetti2007:PRB,zhou2007:NM} For numerical calculations, we diagonalize the tight-binding Hamiltonian,\cite{castroneto2009:RMP,wimmer2010:PRB}
\begin{equation}\label{Model:Hamiltonian}
\begin{aligned}
H = & \sum_i (V_i -\Delta) a_{i}^{\dagger}a_{i} + \sum_i (V_i +\Delta) b_{i}^{\dagger}b_{i} - \\
& -(\sum_{\langle i,j\rangle} t_{ij} a_{i}^{\dagger} b_{j} 
+\sum_{\langle\langle i,j\rangle\rangle} t'_{ij} (a_{i}^{\dagger} a_{j} + b_{i}^{\dagger} b_{j}) + \text{H.c.}),
\end{aligned}
\end{equation}
with the annihilation (creation) operators $a^{(\dagger)}$ and $b^{(\dagger)}$ for the sublattices A and B. The parameter $t_{ij}$ describes the nearest, and $t'_{ij}$ the next-nearest neighbor hopping. The on-site energies account for a position-dependent electrostatic potential $V$ and the mass term $\Delta$. A magnetic field is included using the Peierls phase\cite{peierls1933:ZP} induced by the vector potential ${\bf A}$ through $t_{kl} = t \exp[(i e/\hbar) \int_{{\bf r}_k}^{{\bf r}_l} {\bf A} \cdot  {\rm d} {\bf r} ]$, and similarly for $t^\prime$.

\begin{figure}
 \centering
 \includegraphics[width=0.70\linewidth]{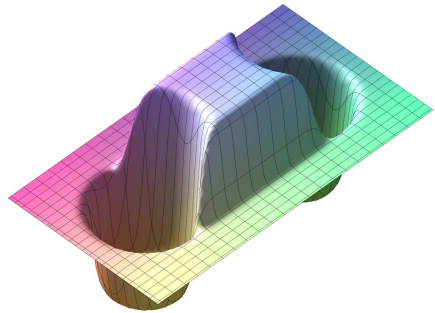}
\caption{(Color online) Illustrative plot of the potential landscape $V(\boldsymbol{r})$, Eq.\,\eqref{Model:Potential}, for parameters $V_0 = -V_1$, and $w=2R$.}
 \label{fig:dotpotential}
\end{figure}

For the dot potential $V$, we use a barbell-like shape with circular disks, aligned along the $x$ axis. The potential profile is defined by [$\boldsymbol{r} = (x,y)$]
\begin{equation}\label{Model:Potential}
V(\boldsymbol{r}) = V_0 \xi_{\text{dot}}(\boldsymbol{r}) + V_1 \max\{|\xi_{\text{bar}}(\boldsymbol{r})| - |\xi_{\text{dot}}(\boldsymbol{r})|,0\} ,
\end{equation}
with the left and right dot defined by 
\begin{equation}\label{Model:2DotPotential}
\xi_{\text{dot}}(\boldsymbol{r}) =  \max\{\xi_{l}(\boldsymbol{r}),\xi_{r}(\boldsymbol{r})\},
\end{equation}
where
\begin{equation}\label{Model:DotPotential}
\xi_{l,r}(\boldsymbol{r}) =  \{1+\exp[(r_{l,r}-R)/\beta]\}^{-1}.
\end{equation}
This represents two single dots with radius $R$, positioned at $\boldsymbol{r}_{l,r} = \boldsymbol{r} \pm (d,0)$, so that $2d$ is the inter-dot distance. The parameter $\beta$ is introduced to smoothen potential edges. The inter-dot barrier of width $w$ is described by 
\begin{equation}\label{Model:BridgePotential}
\begin{aligned}
\xi_{\text{bar}}(\boldsymbol{r}) = & \{1+\exp[(-w/2-y)/\beta])\}^{-1} \times \\
\times & \{1+\exp[(y-w/2)/\beta]\}^{-1} \times \\
\times & \{1+\exp[(-d-x)/\beta]\}^{-1} \times \\
\times & \{1+\exp[(x-d)/\beta]\}^{-1}.
\end{aligned}
\end{equation}
 The geometry is shown in Fig.\,\ref{fig:dotsetup}, and the potential $V(\boldsymbol{r})$, Eq.\,\eqref{Model:Potential}, is sketched in Fig.\,\ref{fig:dotpotential}. The potential $V_0$ represents the single dot confinement depth, while $V_1$ sets the barrier height. In experiments we envisage they are controllable individually by a corresponding local metallic gate, tuning them with respect to the global chemical potential $\mu$. For the following discussion we assume the graphene is undoped, so that $\mu$ is in the middle of the gap $2\Delta$. We measure all energies from there ($\mu=0$).

In the numerics, we diagonalize the Hamiltonian [Eq.~\eqref{Model:Hamiltonian}] with typically 95000 carbon atoms. The grid is extended by $\delta = 25$ nm around the potential barbell to ensure convergence. The finite size of the grid in our numerical calculation leads to edge states at the boundary \cite{akhmerov2008:PRB, wimmer2010:PRB} that would not exist in an extended graphene sheet. To remove these artificial edge states from the calculated spectra, we introduce an additional mass term of $0.3\Delta$ at the grid boundary (sketched by the black line in Fig.\,\ref{fig:dotsetup}).  This trick improves the diagonalization times drastically, without any influence (as we checked) on the confined states we discuss here.

For the quantitative analysis, we use the following material parameters. The hoppings are  $t = 3090$ meV, and $t' = 0.05t$. The graphene lattice constant is 0.246 nm, and the mass is taken as $\Delta = 30$ meV, according to Ref.~\onlinecite{giovannetti2007:PRB}. Unless stated otherwise, we choose the structure parameters as follows: $R = 25$ nm, $d = 55$ nm, and $V_0 = -60$ meV. The thickness of the barrier is $w = 2R$, with a potential strength $V_1$ to be specified. For $V_1 = 0$ the barrier area potential is aligned with the bulk. The smoothening parameter is $\beta = 0.01R$.

\section{Results: single dot}\label{results:singledot}
\begin{figure}
 \centering
 \includegraphics[width=0.90\linewidth]{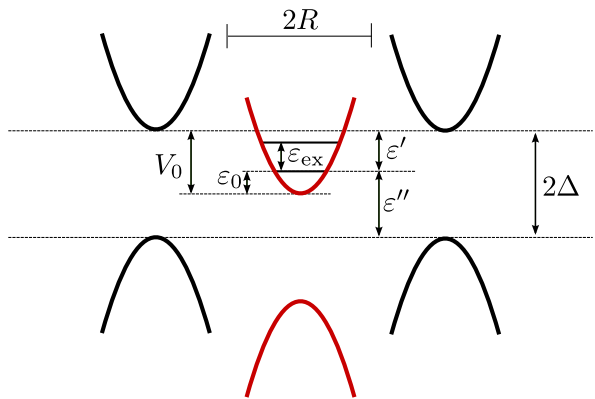}
\caption{(Color online) Schematic energy diagram of a single dot. Within the diameter of $2R$, the energy bands are shifted by the confinement depth $V_0$ (this figure corresponds to $V_0<0$). The ground state is offset from the band bottom by $\epsilon_0$ (ground state energy), from the nearest localized state by $\epsilon_{\rm ex}$ (excitation energy) and from the nearest extended state by $\varepsilon_{\text{ion}} = \min (\varepsilon^{\prime},\varepsilon^{\prime\prime})$ (ionization energy).}
 \label{fig:schematic_energies_SD}
\end{figure}
First, we show how the confinement is created. We start with the example of a single dot, that is $d=0$, $V_1=0$ and consider only a variable depth $V_0$. If $V_0 = 0$, the system is a sheet of graphene with a constant energy gap of $2\Delta=60$\,meV. States with energies $|E|>\Delta$, are extended states of bulk graphene. Consider now $V_0$, is decreased. The conduction band is locally shifted towards lower energies, as shown in Fig.~\ref{fig:schematic_energies_SD}, within the area set by the dot radius $R$. Localized states with discrete energies inside the band gap appear.\cite{trauzettel2007:NP} This is analogous to an $n$-type dot in GaAs.\cite{hanson2007:RMP,zwanenburg2013:RMP}

To describe the localized states spectrum, we define a set of characteristic energies, depicted in Fig.~\ref{fig:schematic_energies_SD}. The ground-state energy $\varepsilon_0$ is the energy difference between $V_0$ and the lowest localized state. The latter is separated from the first excited state by the excitation energy $\varepsilon_{\text{ex}}$. The ionization energy, which we define as the minimal energy required for an excitation from the localized ground state to an extended state, is $\varepsilon_{\text{ion}} = \min (\varepsilon^{\prime},\varepsilon^{\prime\prime})$, where $\varepsilon^{\prime}$ and $\varepsilon^{\prime\prime}$ are the energy differences between the localized ground state and the conduction and valence band edge, respectively. 

\begin{figure}
 \centering
 \includegraphics[width=0.99\linewidth]{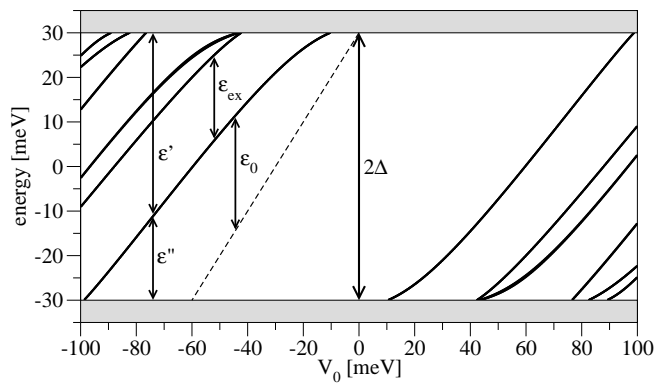}
\caption{(Color online) Calculated energy spectrum of a single dot as a function of the confinement depth $V_0$. For $V_0>0$ ($V_0<0)$, the holes (electrons) are localized in the dot. The gray regions represent extended states. The characteristic energies are indicated.}
 \label{fig:singledot}
\end{figure}

Figure \ref{fig:singledot} shows the calculated energy spectrum of a single dot as a function of the confinement depth $V_0$.  The dot is of $n$ type for $V_0<0$, and $p$ type for $V_0>0$, since the localized state is built from the lowest states of one (conduction) band, and the highest states of another (valence) band, respectively. Neglecting the small next-nearest neighbor hopping $t^\prime$, the Hamiltonian [Eq.~\eqref{Model:Hamiltonian}] has exact symmetry $H(V,B) \to -H(-V,-B)$ upon replacement $(a_i,b_i) \to (b_i, -a_i)$. This translates into a one to one correspondence between low-energy hole-like dot states at $|V_0|$ and electron-like states at $-|V_0|$. From now on we therefore arbitrarily fix $V_0$ negative. There is another, quasi time-reversal, symmetry which connects states at the $K$ valley with ones at $K^\prime$ with opposite total angular momentum $J_z$.\cite{beenakker2008:RMP} As a consequence, in zero magnetic field all states are two-fold degenerate. In general, this degeneracy is a serious obstacle for quantum computation, as it impairs control over the exchange energy.\cite{culcer2009:PRB,zwanenburg2013:RMP} Fortunately, in graphene this degeneracy is split by a finite magnetic field (see below), which couples to the orbital momentum $J_z$.\cite{recher2009:PRB} This is in contrast to the silicon-based dots, where such a direct control knob for the valley splitting does not exist.

\begin{figure}
 \centering
 \includegraphics[width=0.99\linewidth]{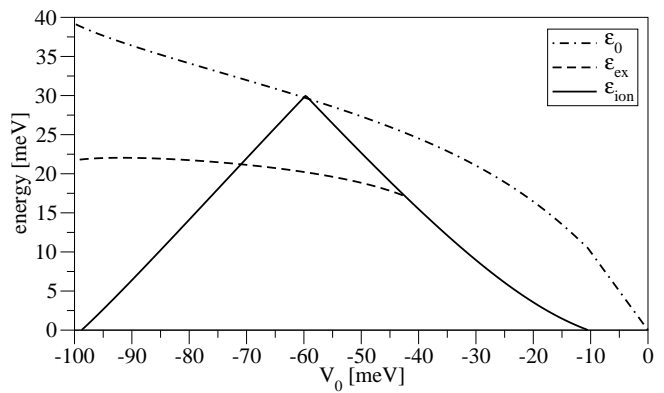}
\caption{Relative ground-state energy $\varepsilon_0$ (dash-dot line), excitation energy $\varepsilon_{\text{ex}}$ (dashed line), and ionization energy $\varepsilon_{\text{ion}}$ (dotted line) as a function of the confinement depth $V_0$.}
 \label{fig:singledotenergies}
\end{figure}

\begin{figure}
 \centering
 \includegraphics[width=0.99\linewidth]{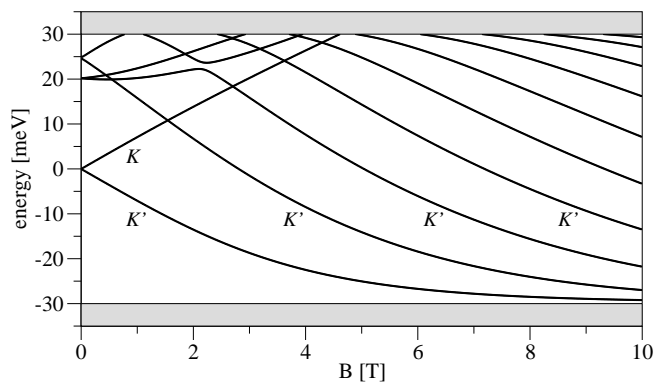}
\caption{Calculated energy spectrum of a single dot ($V_0=-60$\,meV) as a function of a perpendicular magnetic field. The valley index of the lowest states is indicated by $K/K'$. }
 \label{fig:singledotvsB}
\end{figure}

Using the data from Fig.~\ref{fig:singledot}, we plot the characteristic energies in Fig.~\ref{fig:singledotenergies}. We observe that the relative ground-state energy $\varepsilon_0$ increases non-linearly as the dot potential decreases, while the excitation energy $\varepsilon_{\text{ex}}$, which is in our example only defined for $V_0\lesssim -42.5$ meV, is roughly constant. The ionization energy $\varepsilon_{\text{ion}}$ peaks at about $V_0=-60$\,meV, where the ground state is exactly in the middle of the gap. Since we adopt a cylindrical potential shape in our model, the spectrum we obtain is rather different to the one usually met in semiconductor quantum dots with a harmonic confinement. There the energy levels are equidistant, and the ground state energy is half of the excitation energy.\cite{chakraborty,Li2010:PRB} Also, here the lowest localized state hits the bulk band at about $V_0 = -100$ meV. The localized state then necessarily becomes occupied in undoped graphene and may easily hybridize with extended states. This is another difference to semiconductor quantum dots, where the equivalent energy scale is given by the band gap, which is of the order of a few eV.\cite{yu}

The energy spectrum as a function of a perpendicular magnetic field is shown in Fig.~\ref{fig:singledotvsB}. The magnetic field lifts the valley degeneracy by breaking the time reversal symmetry.\cite{recher2009:PRB} Landau levels tend to form.\cite{chen2007:PRL} For our dot parameters the magnetic length $l_{\text{B}}=\sqrt{\hbar/(eB)}$ is of the order of the dot radius $R$ for reasonable magnetic fields.\cite{schnez2008:PRB} More precisely, $l_{\text{B}}$ is 81, 26, and 8 nm for a magnetic field of 0.1, 1, and 10 T, respectively, to be compared to a dot radius of $R=25$ nm. The ``zero-mode'' Landau level\cite{haldane198:PRL,recher2009:PRB} forms at the valence band edge ($-30$ meV) from $K'$ valley states.

\section{Results: double dot}\label{results:doubledot}
\begin{figure}
 \centering
 \includegraphics[width=0.90\linewidth]{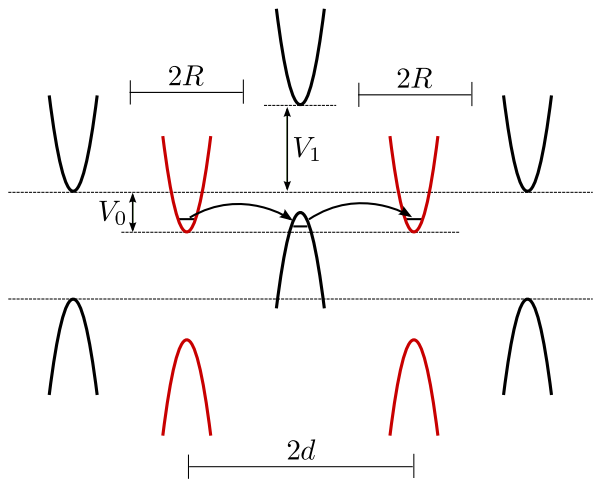}
\caption{(Color online) Schematic energy diagram of a double dot. The energy is shifted locally by $V_0$ for the dots, and by $V_1$ for the barrier (this figure corresponds to $V_0<0<V_1$). The interband tunneling from the left to the right dot via the barrier is indicated by black arrows.}
 \label{fig:schematic_energies_DD}
\end{figure}
\begin{figure}
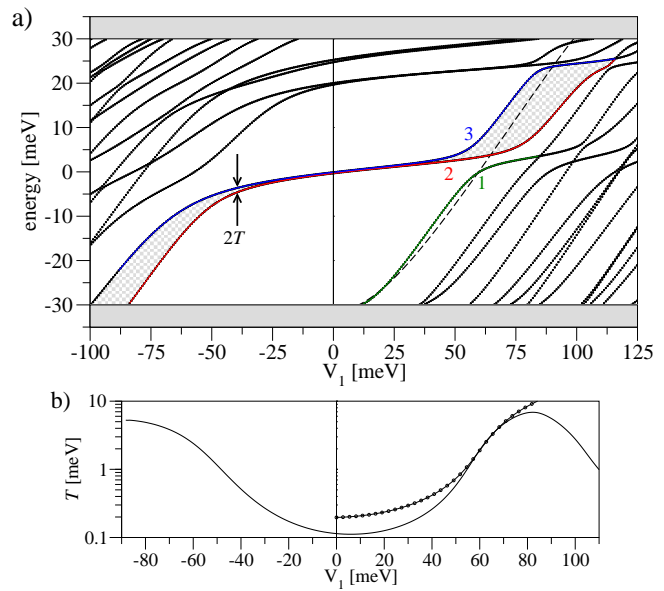

 \centering
   \includegraphics[width=0.99\linewidth]{./DD--vs_V1_dis55nm_Bis0.eps}\\
  \includegraphics[width=0.87\linewidth, trim = 0 0 -20 0]{./DD-tunneling--vs_V1.eps}
\caption{(Color online) (a) Calculated energy spectrum of a double dot ($2d=110$\,nm, $V_0=-60$\,meV) as a function of the barrier potential $V_1$. The red (2), and blue (3) lines denote the dot ground state, and the first excited state, respectively. Half of their energy difference is the tunneling energy $T$.  The shaded region between those two states and the vertical line at $V_1=0$ are a guide for the eye. The green curve (1) denotes the highest localized state of the barrier. The dashed line is the top line for $V_0>0$ from Fig.~\ref{fig:singledot}.  (b) The tunneling energy replotted from the data in (a). Line with symbols is a fit from the model described below Eq.~\eqref{eq:threelevelHamilt}.}
 \label{fig:DD55nm}
\end{figure}

In this section, we study the localized states in a double dot configuration.\cite{moriyama2009:NL,liu2010:NL,molitor2010:EPL,wang2012:AIP} The dot schematic energy diagram is shown in Fig.~\ref{fig:schematic_energies_DD}. The corresponding spectrum, varying the barrier height while keeping the dots confinement depths fixed ($V_0=-60$ meV), is in Fig.~\ref{fig:DD55nm}. Take first the $V_1=0$ point. Here the states in the left and right dots are well isolated by the barrier. The small remaining coupling splits the system eigenstates into symmetric and antisymmetric combinations of the single dot states. They are labeled as 2 and 3 in Fig.~\ref{fig:DD55nm}, and we define the tunneling energy as twice their energy difference.\footnote{We note that for configurations where $V_1<V_0$ there is no barrier in the middle of the dot, and thus the transport of an electron between the single dots happens ballistically, rather than by tunneling. However, as there is no ambiguity in this energy definition, we stick to the name "tunneling energy" throughout for simplicity.} The latter is one possible characterization of the inter-dot coupling\cite{baruffa2010:PRL} and in further we investigate the degree of control over it by electric and magnetic fields. 

\subsection{$nn^\prime n$ coupling regime}

Consider first lowering the barrier between the dots by making $V_1$ negative. The tunneling energy increases due to increasing quantum mechanical overlap of isolated dot eigenstates. In the same time, the energy distance to the next orbital level decreases. This is expected, as at $V_1 \sim V_0=-60$ meV the system resembles more a prolongated single dot, rather than a tunnel coupled double dot. For even larger negative $V_1$ the lowest states are more and more localized in the barrier itself and the system qualitatively changes back to a single dot---the tunneling energy saturates and energy distance to the next level increases. This behavior, referred to as the $nn^\prime n$ coupling,\cite{liu2009:PRB} is analogous to a GaAs dot.

\subsection{Coupling minimum}

Consider now that, starting again from zero, $V_1$ is increased. In GaAs, such an increase of the barrier between the two dots makes the tunneling smaller (exponentially). Such high sensitivity of the inter-dot coupling to electrostatic potential\cite{gorkov2003:PRB,stano2005:PRB} is at the heart of the versatile control of spin qubits in GaAs dots. Looking at Fig.~\ref{fig:DD55nm}(b), however, a decrease in tunneling with $V_1$ is very modest and soon changes into an increase (the tunneling reaches minimum at $V_1=6$ meV, which is 3 $\mu$eV below its value of $110$ $\mu$eV at $V_1=0$). This turn in trend is caused by the presence of the $p$-like (valence) states localized inside the barrier,\cite{trauzettel2007:NP} through which the $n$-like dot states are coupled effectively. This is referred to as the $npn$ coupling.\cite{liu2009:PRB,rozhkov2011:PR}
The arising limit-from-below on the inter-dot coupling, which can be considered as the manifestation of the Klein paradox,\cite{calogeracos1999:CP} then imposes possibly a serious limitation on the extent of control over the double dot states. It is important to understand what parameters set the achievable coupling minimum.

\begin{figure}
 \centering
 \includegraphics[width=0.99\linewidth]{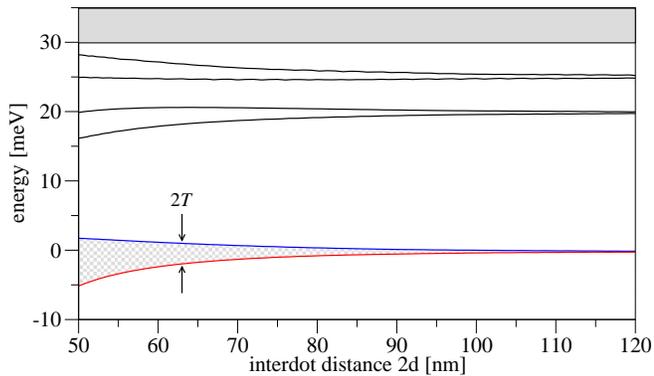}
\caption{(Color online) Calculated energy spectrum of a double dot ($V_0=-60$\,meV, $V_1=0$) as a function of the interdot distance $2d$. The red and blue curve give the localized ground and first excited state, respectively. The tunneling energy $T$ is indicated.}
 \label{fig:vs2d}
\end{figure}

\begin{figure}
 \centering
 \includegraphics[width=0.99\linewidth]{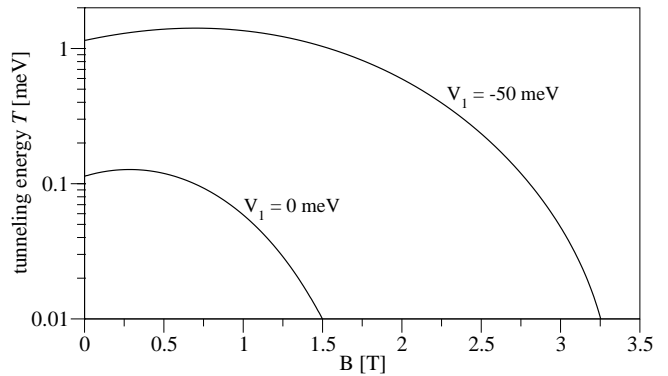}
\caption{Calculated tunneling energy $T$ for a graphene double dot ($2d = 110$\,nm, $V_0=-60$\,meV) as a function of perpendicular magnetic field for two values of the barrier potential.}
 \label{fig:tunneling_vsB}
\end{figure}

We expect the minimum to occur, very roughly, when the dot confined state is energetically half-way between the barrier conduction and valence bands. For our parameter $V_0=-60$ meV, resulting in $\epsilon_0=30$ meV, this estimate gives the minimum at $V_1=0$, which is close to the value observed in numerics.
The value of the minimal tunneling will decrease with the barrier height and length. The former is proportional to the induced mass $\Delta$, while the latter is given by the inter-dot distance. This is confirmed in Fig.~\ref{fig:vs2d}, where a drop of tunneling with the  inter-dot distance is demonstrated. Alternatively to changing the electrostatic barrier, the tunneling can be suppressed by a perpendicular magnetic field.\footnote{An inhomogeneous magnetic field can even provide for a complete confinement, see Ref.~\onlinecite{demartino2007:PRL}.}
 Localization of electrons by its orbital effects, demonstrated by the tunneling suppression, was studied in detail in GaAs\cite{stano2005:PRB} and silicon\cite{raith2011:PRB} quantum dots. Our expectation of an analogous behavior in graphene is confirmed by numerical results, shown in Fig.~\ref{fig:tunneling_vsB}. There it is shown that the perpendicular field of few Tesla suppresses the tunneling by 1-2 orders of magnitude, depending on the barrier potential.

\subsection{$npn$ coupling}

\begin{figure}
 \centering
 \includegraphics[width=0.99\linewidth]{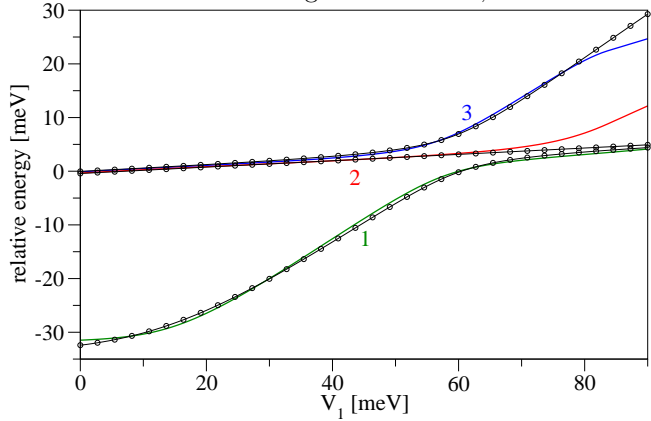}
\caption{(Color online) Eigenenergies of the states 1, 2, and 3, taken from Fig.~\ref{fig:DD55nm}, (solid lines) and from the analytical model, described below Eq.~\eqref{eq:threelevelHamilt}, (lines with circles). A linear trend, $\delta \epsilon = 0.0593 V_1$, fitted from the exact values of the level 2 energy, was added to all fitted values.}
 \label{fig:fitting}
\end{figure}

Let us finish the analysis of the effects of the barrier potential $V_1$, now considering the $npn$ coupling regime. It is signaled by a change in the trend of the inter-dot coupling, which starts to increase upon increasing the barrier height. In Fig.~\ref{fig:DD55nm} this happens at $V_1\approx 6$ meV. Beyond this point, the left and right single dots are coupled most effectively through the $p$-like localized state formed in the barrier, tagged as line 1 in Fig.~\ref{fig:DD55nm}(a). In this regime, the double dot low energy spectrum is well characterized considering three states (the left dot, the right dot, and the barrier), with the neighboring pairs coupled by a tunneling matrix element $\tau$. To demonstrate this, we replot the three energies from Fig.~\ref{fig:DD55nm} into Fig.~\ref{fig:fitting} and fit them with the spectrum of the following Hamiltonian:
\begin{eqnarray}\label{eq:threelevelHamilt}
H = \begin{pmatrix}
0 & \tau & 0\\
\tau & \varepsilon & \tau\\
0 & \tau & 0 \end{pmatrix}.
\end{eqnarray}
First, the energy offset of the $p$-state is well approximated by the single dot ground state energy $\epsilon(V_1) \approx -\Delta+V_1-\epsilon_0(V_1)$. This is shown by plotting the approximation (that is, the top line for $V_0>0$ from Fig.~\ref{fig:singledot}) as a dashed line in Fig.~\ref{fig:DD55nm}(a). We note that this is rather a coincidence, as the effective dot shape of the $p$-region is different from the single dot considered before (though both of these energies should indeed scale similarly, as the areas of the dot and the barrier are comparable in our model). However, for our purpose of extracting the tunneling matrix element $\tau$ we can still make use of this fact and need not fit the energy offset of the $p$-state separately. Second, the tunneling matrix element $\tau$ is obtained from the width of the anti crossing of lines 1 and 3 [at $V_1\approx 60$ meV in Fig.~\ref{fig:DD55nm}(a)], from where we get $\tau\approx 2.5$ meV. With this, basically a single parameter fit, we obtain a very good agreement with the exact numerics, as seen in Fig.~\ref{fig:fitting}, not only close to the anti crossing, but through out the whole $npn$ regime. The model starts to deviate only once higher excited $p$-states in the barrier anti-cross with the $n$-like dot states (beyond $V_1\gtrsim 70$ meV). The correspondence for lower $V_1$ is demonstrated by comparing the exact tunneling energy with the one obtained from our model [Eq.~\eqref{eq:threelevelHamilt}], which we do in Fig.~\ref{fig:DD55nm}b. Since at its minimum the tunneling energy does not differ from the model by more than a factor of 2, we estimate that through out the $npn$ region $\tau$ does not vary more than by a factor of $1/\surd{2}$. This stability of the tunneling matrix element over a very large range of gate voltages is perhaps surprising, but welcomed as it justifies the analysis of the double dot structure in terms of isolated eigenstates of the dots and the barrier.

We finish with a note that the system in the $npn$ regime seems well suited for adiabatic passage protocols.\cite{greentree2004:PRB,fabian2005:PRB,cole2008:PRB} Namely, these are based on a Hamiltonian such as Eq.\eqref{eq:threelevelHamilt}, if one can control the two matrix elements independently. Such a control might be possible using an electric field $E$ applied along the dot axis, which changes the Hamiltonian from Eq.~\eqref{eq:threelevelHamilt} into
\begin{eqnarray}\label{threelevelHamilt2}
H = \begin{pmatrix}
-eEd & \tau^\prime & 0\\
\tau^\prime & \varepsilon & \tau^{\prime\prime}\\
0 & \tau^{\prime\prime} & eEd \end{pmatrix}.
\end{eqnarray}
The important feature is that the electric field shifts the dot and barrier states in opposite directions, as they are $n$- and $p$-like, respectively. This will change the tunneling matrix elements in an asymmetric way, e.g., $\tau^\prime > \tau > \tau^{\prime\prime}$. Based on numerical results (not shown) we find that, unfortunately, in our model the single dot confinement edges are so steep that the electric field induced asymmetry in $\tau^\prime$ is too small compared to the energy shifts $\pm eEd$ and therefore not suitable for adiabatic passage protocols. Namely, the latter require an efficient way to independently tune $\tau^\prime$ and $\tau^{\prime\prime}$, well into the limits $\tau^\prime \gg \tau^{\prime\prime}$ and $\tau^\prime \ll \tau^{\prime\prime}$, and without simultaneously inducing substantial energy offsets between the states. Other potential profiles might be better suited for this task. 

\section{Conclusions}\label{conclusions}
We investigated the energy spectrum of gated single and laterally coupled double quantum dots in gapped graphene. For the single dot, we presented the characteristic energies of the localized states: the ground state energy, the excitation energy, and the ionization energy. We also showed the influence of a perpendicular magnetic field, including the lifting of the valley degeneracy. 

We primarily focused on the inter-dot coupling in a double dot configuration, the control over which is crucial for few electron quantum dot states manipulations. The extent of the electrical control over the electron tunneling in graphene is {\it a priori} not clear, because of the Klein tunneling effect. Our results suggest that these limitations are to a large extent overcome in gapped graphene and the inter-dot tunneling can be varied, for reasonable parameters, by orders of magnitude. The most important difference to a GaAs dot is that in a graphene dot there is a minimum below which the tunneling can not be reduced. This is due to the presence of states localized in the barrier, what can be seen as an artifact of the Klein tunneling. The achievable coupling minimum is set by the dot geometry and the substrate-induced gap. For these parameters fixed, the tunneling can be further reduced by a perpendicular magnetic field. The presence of the states in the barrier, on the other hand, offers new regimes of operation, unaccessible to GaAs material, such as the $npn$ double dot.

\acknowledgments
We would like to thank Jan Bundesmann and Sebastian Lautenschlager for useful discussions. This work was supported by DFG under grant SPP 1285 and SFB 689. P.~S.~acknowledges support from QUTE and COQI APVV-0646-10.

\bibliography{../../references/quantum_dot}

\end{document}